\documentclass[aps,prb,preprint]{revtex4-1}
\usepackage{booktabs}
\usepackage{amsmath,graphicx}
\begin{document}

\title{Direct observation of non-fully-symmetric coherent optical phonons by femtosecond x-ray diffraction}

\author{S. L. Johnson}
\affiliation{Department of Physics, ETH Zurich, CH-8093 Zurich, Switzerland}
\author{P. Beaud}
\affiliation{Swiss Light Source, Paul Scherrer Institut, CH-5232 Villigen PSI, Switzerland}
\author{E. M\"ohr-Vorobeva}
\altaffiliation[Current address:  ]{Physics Department, University of Oxford, Clarendon Laboratory, Parks Road, Oxford, UK}
\affiliation{Swiss Light Source, Paul Scherrer Institut, CH-5232 Villigen PSI, Switzerland}
\author{A. Caviezel}
\author{G. Ingold}
\affiliation{Swiss Light Source, Paul Scherrer Institut, CH-5232 Villigen PSI, Switzerland}

\author{C. J. Milne}
\altaffiliation[Current address:  ]{Paul Scherrer Institut, CH-5232 Villigen PSI, Switzerland}
\affiliation{Laboratoire de Spectroscopie Ultrarapide, Ecole Polytechnique F\'{e}d\'{e}rale de Lausanne, 1015 Lausanne, Switzerland}

\begin{abstract}
We directly measure by femtosecond time-resolved x-ray diffraction the  E$_g$ symmetry coherent phonon excited in bismuth by strong optical excitation.  The magnitude of the E$_g$ mode observed is 0.2~pm peak-to-peak, compared against the  $2.7$~pm initial displacement of the fully-symmetric A$_{1g}$ mode.  The much smaller motion of the E$_g$ mode is a consequence of the short lifetime of the electronic states that drive the atomic motion.  The experimentally measured magnitude of the E$_g$ motion allows us to rule out a previously suggested scenario for explaining the dynamics in bismuth that relies on strong coupling between these modes.
\end{abstract}

\maketitle

\section{Introduction}

Strong optical excitation of solids on time scales much smaller than typical vibrational periods often leads to coherent structural dynamics that is distinct from any dynamics exhibited in equilibrium.  The best known example is that of  elemental bismuth, where the electronic state distribution created by short pulse optical excitation leads to large-amplitude coherent vibrations that cannot be observed under conditions of thermal equilibrium~\cite{SokolowskiTinten:2003ul,Murray:2007ce,Hase:2002in,Fritz:2007wq}. Similar effects have been observed in related materials, such as tellurium~\cite{Hunsche:1995vc,Roeser:2004wk,Johnson:2009dx}.  

Coherent vibrations can in principle mediate phase transitions that involve a change in unit cell structure.  In these cases the period of the vibrational mode sets a lower limit on the speed of the transition.  The simplest example of this is the Peierls charge-density-wave transition in an idealized one-dimensional material with one atom per unit cell and one valence electron per atom~\cite{ErnstPeierls:1991tm}.  At low temperatures, this structure is unstable with respect to the formation of a superlattice distortion with a periodicity of twice the spacing between the atoms.  Sudden electronic excitation of the system in this low-temperature phase results in a coherent oscillation of the atoms as the interatomic potential energy surface suddenly relaxes back toward the unmodulated, high-temperature structure.  This is qualitatively what happens in the above-mentioned real systems like bismuth and tellurium.  At sufficiently high levels of electronic excitation, models and some experiments have indicated that the interatomic potential can relax fully to the normal, unmodulated phase resulting in a fast phase transition~\cite{Cavalleri:2004eh,Murray:2005wf,Lu:2010ts,Beaud:2009br,MohrVorobeva:2011cb}. 

It has, however, long been recognized that short-pulse lasers can also induce large-amplitude coherent optical mode oscillations that break symmetry operations present in the initial state~\cite{Merlin:1997uf}.  These kinds of motion are critical to understand in the context of inducing phase transitions that reduce the symmetry of a crystal.  Bismuth is well-known as a model system for ultrafast structural dynamics, and as such it provides an excellent opportunity to study in detail the dynamics of non-fully-symmetric coherent phonon modes.

\begin{figure}
\includegraphics{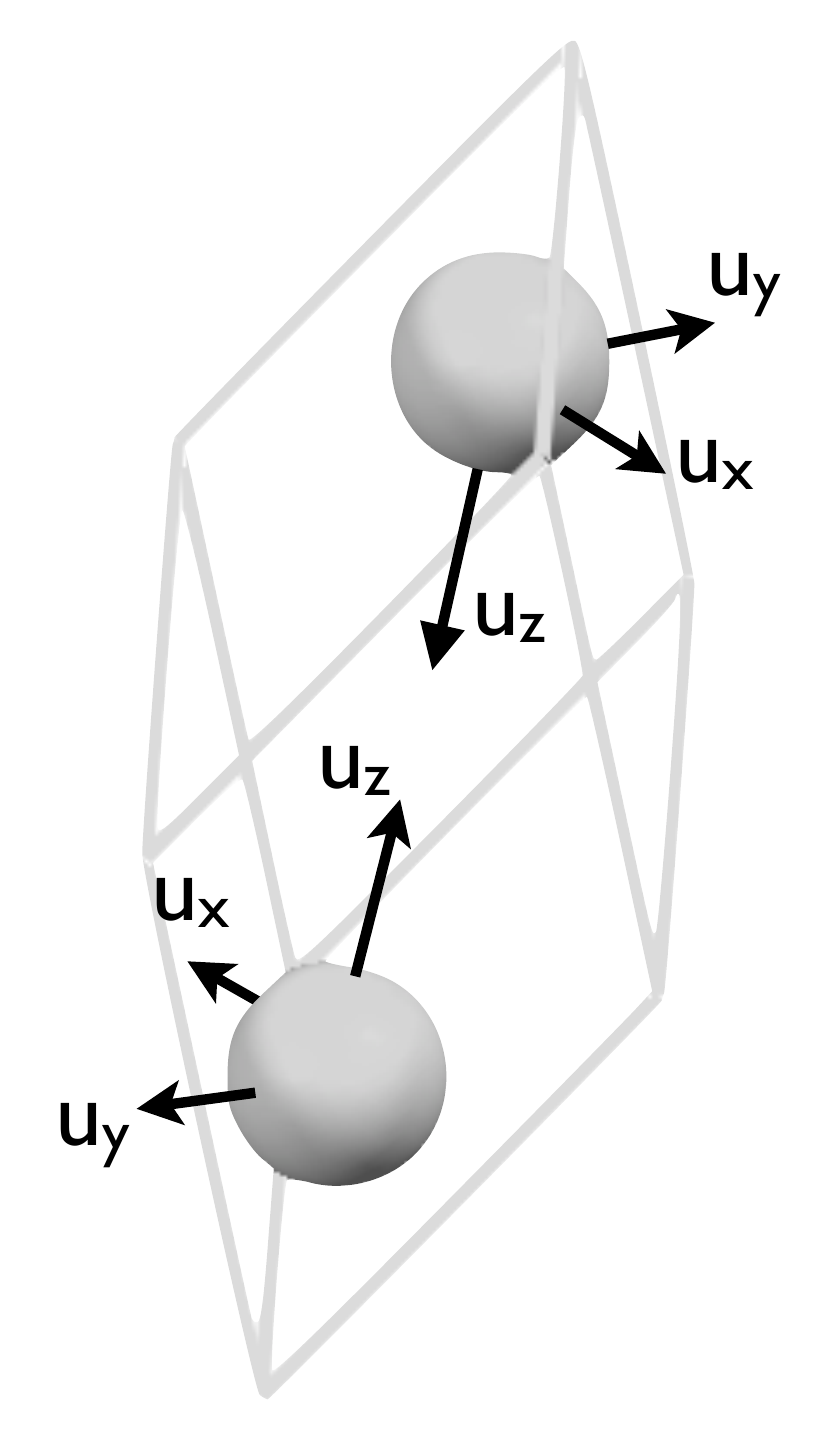}
\caption{Sketch of the rhombohedral unit cell of bismuth, showing the displacements corresponding to each zone-center optical phonon.  The $u_x$ displacement moves the atoms parallel to a $C_2$ symmetry axis, whereas the $u_y$ displacement moves the atoms within a mirror plane containing the $C_3$ symmetry axis.}\label{fig:BiUC} 
\end{figure}

The unit cell structure of bismuth under equilibrium conditions is shown in Fig.~\ref{fig:BiUC}.  The primitive unit cell is rhombohedral with lattice constant $a_r=4.7574\,\textrm{\AA}$ and $\alpha=57.09^\circ$ at room temperature~\cite{Fischer:1978wu}.  In cartesian coordinates, we can represent the primitive cell vectors as
\begin{align}
\mathbf{a}_1 &=  \left(-{1 \over 2} a_h , - {\sqrt{3} \over 6} a_h  , { 1 \over 3} c \right)\label{eq:as}\\ 
\mathbf{a}_2 &=  \left({1 \over 2} a_h , - {\sqrt{3} \over 6} a_h  , { 1 \over 3} c \right)\\ 
\mathbf{a}_3 &=  \left(0,{\sqrt{3} \over 3} a_h , { 1 \over 3} c \right) 
\end{align}
where $a_h = 4.54675\,\textrm{\AA}$ and $c=11.90291\,\textrm{\AA}$ are the unit cell constants of the corresponding non-primitive hexagonal cell. 
The basis consists of two atoms, positioned along the diagonal of the unit cell at $
\left(0,0,\pm \zeta c \right) $, with $\zeta=0.2334$.  

In bismuth there exist three independent optical phonon modes at zero wave vector:  one A$_{1g}$ mode and a pair of degenerate E$_g$ modes.  The A$_{1g}$ mode preserves the symmetry of the crystal and can be described as an oscillatory movement of the two basis atoms in opposite directions along the body diagonal of the unit cell with an amplitude $u_z$.  The eigenvectors for the E$_g$ modes are sketched in Fig.~\ref{fig:BiUC}:  these modes correspond to motion of the basis atoms in equal and opposite directions perpendicular to the C$_3$ axis.  Displacement along any E$_g$ coordinate breaks the C$_3$ symmetry of the cell.  A displacement $u_x$ along the direction of the x-axis will also break all C$_2$ and mirror plane symmetries.  For the orthogonal $u_y$ displacement, exactly one C$_2$ axis and one mirror plane symmetry are preserved.

Impulsive stimulated Raman scattering from an optical pulse with a duration much smaller than the period of these modes can create large amplitude coherent motions of zero wave vector phonon modes~\cite{Merlin:1997uf}.  
Stevens et al.~\cite{Stevens:2002ww} proposed a treatment of this process for absorbing media that involves for each phonon mode two frequency-dependent second-rank tensors.  
One of these tensors that we will call $\pi_{kl}^j$ ($j=x,y,z$) is of interest here since it gives the force on an atom along the $u_j$ directions:
\begin{equation}
F_j(t) = {\epsilon_0 v_c \over 8 \pi} \sum_{kl} \int_{-\infty}^{\infty} \int_{-\infty}^{\infty} e^{-i \Omega t} E_k(\omega) \pi_{kl}^j(\omega, \omega-\Omega) E_l^*(\omega-\Omega)  d\omega d\Omega  \label{eq:F} + \textrm{c.c.}
\end{equation}
where  $E_k(\omega)$ is the Fourier transform of the optical electric field, and $v_c$ is the volume of a unit cell~\footnote{Note that here we define $\pi_{kl}^j$ with different constants than does Ref.~\onlinecite{Stevens:2002ww}, so as to make $\pi_{kl}^j$ independent of crystal size and Eq.~\ref{eq:F} simpler.}.
Each phonon mode is associated with a different tensor $\pi_{kl}^j$.  
The symmetry properties of the phonon modes place restrictions on the form of $\pi_{kl}^j$.  For the $A_{1g}$ mode the symmetry of the crystal is preserved for any value of the displacement $u_z$, so the tensor takes the diagonal form
\begin{equation}
\pi_{kl}^{z} = \left( 
\begin{array}{ccc} 
a & 0 & 0 \\
0 & a & 0 \\
0 & 0 & b \\
\end{array}
\right).\label{eq:piz}
\end{equation}
For the E$_g$ modes, the $C_3$ symmetry implies there are two tensors
\begin{equation}
\pi_{kl}^{x} = \left( 
\begin{array}{ccc} 
0 & -d & -f \\
-d & 0 & 0 \\
-f & 0 & 0 \\
\end{array}
\right)\label{eq:pix}
\end{equation}
\begin{equation}
\pi_{kl}^{y} = \left( 
\begin{array}{ccc} 
d & 0 & 0 \\
0 & -d & f \\
0 & f & 0 \\
\end{array}
\right).\label{eq:piy}
\end{equation}
If the optical photon energy is far from electronic resonances the values of $d$ and $f$ are in principle measurable from spontaneous Raman scattering.  Near resonances the situation becomes more complex since the dynamics of the polarization density rely on another non-equivalent tensor with the same symmetry form but different frequency dependence~\cite{Stevens:2002ww}.

Coherent excitation of the $A_{1g}$ and $E_g$ modes in bismuth has been previously observed with pump-probe optical reflectivity experiments~\cite{Hase:2002in,Misochko:2006gd,Li:2011vb}.  One of the more intriguing outcomes of this work is the suggestion of coupling between the two modes~\cite{Hase:2002in,Zijlstra:2006wt}.  The strength of this coupling is highly sensitive to the amplitude of the coherent E$_g$ mode.  Consequently, evaluating possible coupling mechanisms requires some way to accurately experimentally measure the actual atomic displacements of both the E$_g$ and A$_{1g}$ modes, quantities that are not available from transient reflectivity measurements.  More generally, quantitative information on the magnitude of non-fully-symmetric modes permits investigation into the possibility of inducing structural phase transitions where the target has a symmetry lower than the initial structure.
This requires a more direct type of measurement, now possible using femtosecond time-resolved diffraction techniques~\cite{Johnson:2010em,SokolowskiTinten:2003ul,Fritz:2007wq}.

\section{Experiment}

\begin{figure}
\includegraphics[width=15cm]{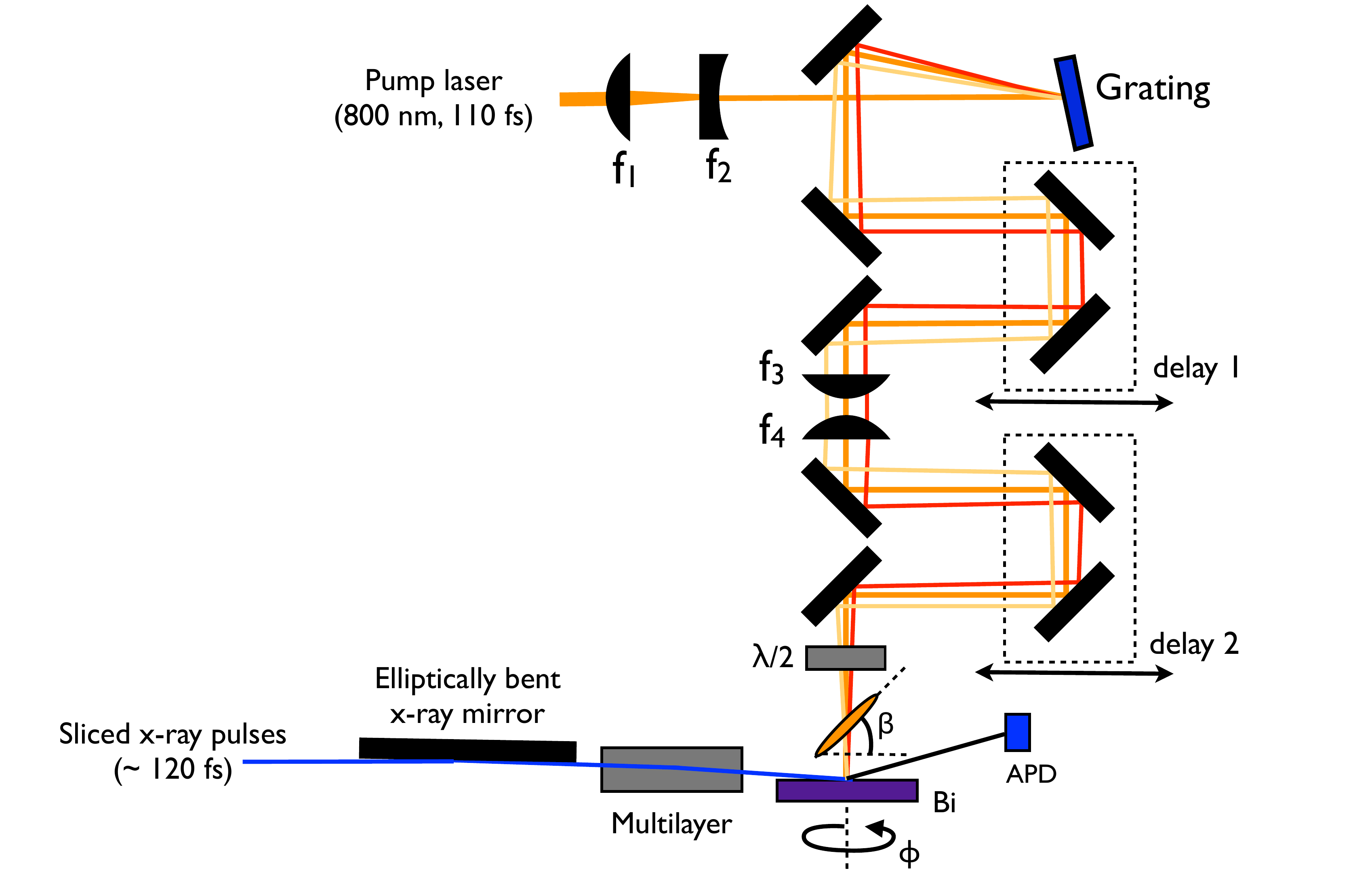}
\caption{Simplified sketch of the experimental setup.  To excite the dynamics in the sample, a femtosecond pulse from a commercial regenerative amplifier is tilted by imaging the first order reflection from a grating onto the sample.  The polarization of the pump is controlled by a $\lambda/2$ plate inserted just before the sample.  To probe the dynamics, the sliced femtosecond duration x-rays reflect from an elliptically bent mirror vertically, and then horizontally from a multilayer mirror before encountering the sample.  Diffracted x-rays are then detected using an avalanche photodiode.  The reflection conditions for different diffraction peaks are realized by adjusting the sample rotation $\phi$.}\label{fig:sketch}
\end{figure}

Figure~\ref{fig:sketch} shows a conceptual sketch of the femtosecond x-ray diffraction measurements.  An intense infrared pump pulse (800~nm, 115~fs, 1~kHz) generated by an amplified femtosecond laser system excites the sample.  To probe the changes in structure, we use the 140~fs femtosecond duration pulses generated by the electron-beam slicing method at the Swiss Light Source~\cite{Beaud:2007fz}.  As in our previous investigations on bismuth~\cite{Johnson:2008ku,Johnson:2009ty,Johnson:2010em}, a toroidally bent mirror first collimates the beam vertically and focuses it horizontally to achieve a spot size of 250 $\mu$m at the sample.  Further downstream, an elliptically bent mirror focuses the beam vertically to a size of 7~$\mu$m at the sample position.  Before hitting the sample, the x-rays reflect horizontally from a Mo/B$_4$C multilayer that weakly monochromatizes the x-ray beam to 7.05 keV with a bandwidth of 1.2\%.  
The x-rays enter the sample at a grazing angle of $0.5^\circ$ in order to limit the x-ray field penetration depth to 30~nm and to better match the pump excitation depth.     The surface of the sample was miscut from the $(111)$ plane by $\alpha_m = 57^\circ$ toward the $(2\,\bar{1}\,\bar{1})$
 plane. Rotating the sample about its surface normal allows access to both the $(1\,1\,1)$ and $(0\,0\,1)$ x-ray reflections.

Previous femtosecond x-ray diffraction experiments on bismuth have used a pump beam with a large incidence angle in an effort to minimize the angle between the pump beam and the x-rays.  The non-zero size of the x-ray beam projected onto the crystal surface causes a loss in effective time resolution from the difference in relative arrival times between the pump and x-ray beams at different points along the sample.  For a $7$~$\mu$m vertical focus size of the x-rays at a grazing angle of $0.5^\circ$, the time resolution from this geometrical factor runs from $40$~fs at $80^\circ$ pump incidence angle ($10^\circ$ grazing) to $2.6$~ps at normal pump incidence.  High incidence angles for the pump beam come, however, with a price:  the transmission and reflectivity, especially near the Brewster angle, become highly polarization dependent.  In the case of bismuth at $80^\circ$ incidence, the reflectivity of $800$~nm light changes from 33\% for p-polarization to 94\% for s-polarization.  This makes it difficult to study the effects of changes in polarization alone on the dynamics, which is one of the key differences between the excitation of the A$_{1g}$ and E$_g$ modes.  For this it is advantageous to make the pump beam arrive at normal incidence where the difference in transmission and reflectivity for different polarizations is minimized.  
  
To achieve this without compromising time resolution, we adopted a scheme to tilt the  intensity front of the pump laser pulse by $45^\circ$ to make the pump-probe delay time nearly constant for normal incidence across the entire pumped region of the crystal.  Practically, this is done by imaging the first order reflection from a grating onto the sample~\cite{Hebling:1996vr}.  The tilt angle $\beta$ is set according to
\begin{equation}
\tan(\beta) = { \lambda_0 \over  d_g M \sqrt{ 1-\left({\lambda_0 / d_g} - \sin\Theta\right)^2}  } 
\end{equation}
where $\lambda_0$ is the center wavelength of the laser pulse, $d_g$ is the grating line spacing, $\Theta$ is the grating incidence angle, and $M$ is the magnification of the imaging system.  Both $\Theta$ and $M$ may be adjusted to tune $\beta$.  In our case we user a blazed gold-coated grating (Spectrogon PC0900) with $d_g = 900$~lines/mm and $\Theta = 17^\circ$, and so to achieve $\beta = 45^\circ$ we adjusted the path delays to make $M = 0.80$ while maintaining the image point at the sample.  The imaging was performed with a pair of plano-convex lenses (focal lengths $f_3 = 1.5\,$m and $f_4 = 1\,$m) configured to compensate partially for spherical aberration~\cite{Kidger:2002tl}.  The alignment and tilt angle was verified prior to the x-ray experiment by cross-correlating an untilted pick-off of the 800 nm beam with the tilted beam using a nonlinear BBO crystal cut for phase-matched second harmonic generation at the optical position of the sample.    The size of the laser spot on the grating was controlled by a pair of lenses $f_1$ and $f_2$ that reduce the size of the beam to a diameter of 3~mm (FWHM).  The incident fluence on the sample was set to 6 mJ/cm$^2$, corresponding to 1.6 mJ/cm$^2$ absorbed fluence.  

\section{Results}

Because the $(111)$ atomic planes lie perpendicular to the diagonal of the unit cell, x-ray diffraction from these planes is insensitive to atomic motion along the $u_x$ or $u_y$ coordinates.  The intensity of this reflection is, however, strongly sensitive to coherent motion of the $u_z$ coordinate~\cite{SokolowskiTinten:2003ul,Fritz:2007wq,Beaud:2007fz, Johnson:2008ku}.  The $(0\,0\,1)$ planes lie at an angle of $71.6^\circ$ from the $(111)$ planes and are strongly sensitive to both the A$_{1g}$ mode and to the E$_g$ mode with displacement along the $u_y$ coordinate.  Quantitatively, the intensity $I$ of a diffraction peak with reciprocal lattice vector $\mathbf{G}$ in a kinematic approximation is proportional to $|F|^2$, where $F = \sum_j f_j e^{i\mathbf{G} \cdot \mathbf{r}_j}$ is the structure factor.  Here the sum is over all basis atoms with index $j$, $f_j$ are the atomic form factors, and $\mathbf{r}_j$ are the basis atom positions.  
For the specific case of bismuth and the $(111)$ and $(0 0 1)$ reflections we may write
\begin{equation}
{I_{111} \over I_{111}^0} = \left|{F_{111} \over F_{111}^0}\right|^2 = {\cos^2( 6 \pi (\zeta + u_z / c)) \over \cos^2( 6 \pi \zeta)}\label{eq:I111}
\end{equation}
and
\begin{equation}
{I_{001} \over I_{00{1}}^0} = \left| {F_{00{1}} \over F_{00{1}}^0}\right|^2 = {\cos^2( 2 \pi ( {2 \sqrt{3} \over 3 a_h} u_y + (\zeta+u_z/c) )) \over \cos^2( 2 \pi \zeta)}\label{eq:I001}
\end{equation}
where $I_{111}^0$ and $I_{00{1}}^0$ are the measured intensities in equilibrium ($u_x=u_y=u_z=0$).  Here we neglect treatment of changes in the Debye-Waller factor that, while present in bismuth~\cite{Johnson:2009ty}, lead to changes of less than 1\% for the diffraction peaks we study here.

\begin{figure}
\includegraphics[width=15cm]{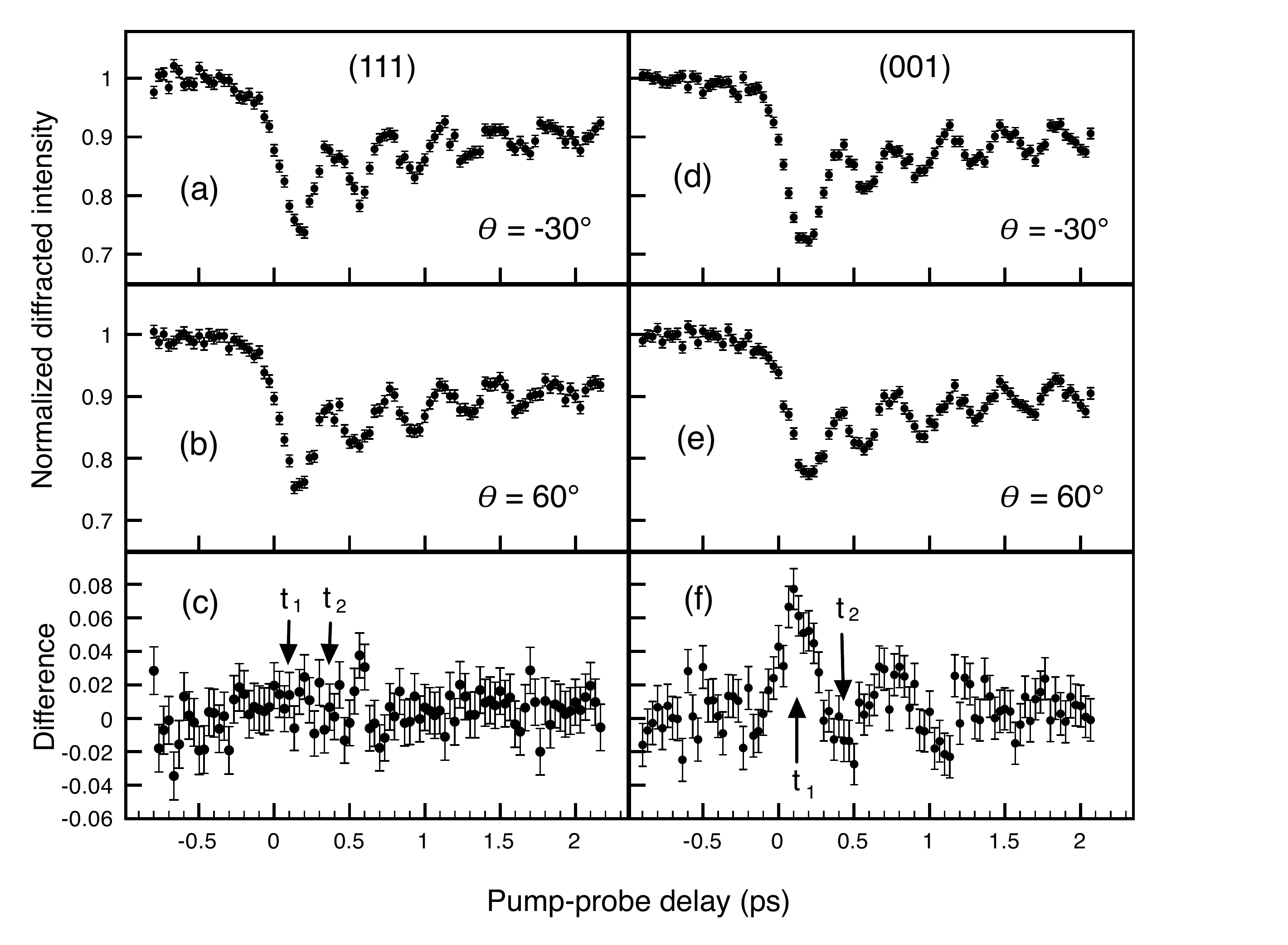}
\caption{Summary of time-resolved diffraction data from a bismuth single crystal at a temperature of 170~K.  The left panels show data for the $(111)$ reflection, whereas the right panels show data for the $(100)$ reflection.  The topmost panels show the data for a pump polarization $\theta = -30^\circ$;  the middle panels show the data for $\theta = 60^\circ$.  The bottom panels show the difference in the diffracted intensity between the two polarizations.  Whereas the $(111)$ peak shows no significant difference between the polarizations, the $(001)$ data shows evidence of a strongly damped oscillation.  The arrows indicate the times that are used in Fig.~\ref{fig:poldep} to study the polarization dependence in more detail.
}\label{fig:diffpanel}
\end{figure}

Figure~\ref{fig:diffpanel} shows the diffracted intensity from the $(111)$ and $(00{1})$ diffraction peaks as a function of pump-probe delay at an initial sample temperature of $170\,\textrm{K}$.  For these measurements a half-wave plate controlled the polarization $\theta$ of the pump with respect to the projection of the $C_3$ axis with an uncertainty of $\pm 2^\circ$.  Positive values of $
\theta$ denote counter-clockwise rotation.  In Figs.~\ref{fig:diffpanel}(a) and \ref{fig:diffpanel}(d), $\theta = -30^\circ$.  In Figs.~\ref{fig:diffpanel}(b) and \ref{fig:diffpanel}(e), this was changed to $\theta = 60^\circ$.  There is little difference in the time-dependence of the $(111)$ diffraction, but the dynamics from the $(001)$ peak show noticeable changes.  
This becomes clear in Figs.~\ref{fig:diffpanel}(c) and \ref{fig:diffpanel}(f) which show the difference between the data sets when changing only the pump polarization.  
For the $(111)$ diffraction the difference is zero within the errors, while for the $(001)$ diffraction there is a clear oscillation starting at a pump-probe delay of zero.  
To further establish that these oscillations depend sensitively on the polarization, Fig.~\ref{fig:poldep} shows for each diffraction peak the polarization dependence of the ratio of the pump-induced intensity change $\Delta I(t_1)/\Delta I(t_2)$, where $t_1=80\,\textrm{fs}$ and $t_2=370\,\textrm{fs}$ are pump-probe delays corresponding to the approximate times of the first two extrema of the oscillation.
For the $(111)$ peak this is independent of polarization within the experimental errors, whereas for the $(001)$ peak there is a strong dependence with a period of $180^\circ$.  

\section{Discussion}

A quantitative analysis of the polarization dependence requires us to evaluate the optical field inside the crystal, which is in general straightforward but algebraically complicated due to birefringence of the crystal.  Fortunately, at 800 nm in bismuth the birefringence is fairly weak:  $\epsilon_{11}= \epsilon_{22} = -18.4+ 28i$ and $\epsilon_{33} = -13.5 + 28 i$~\cite{Lenham:1965to}.  We accordingly make the approximation of an isotropic dielectric tensor with $\epsilon \approx \sum_j \epsilon_{jj} / 3 =  -16.8+28i$.  
If $E_0(t)$ is the amplitude of the incident electric field, inside the crystal we then have 
\begin{equation}
\mathbf{E}(t) = {2 E_0(t) \over 1 + \sqrt{\epsilon}} \left(
\begin{array}{c}
{ \sqrt{3} \over 2 } \cos \alpha_m \cos \theta - {1 \over 2} \sin \theta \\
{ 1 \over 2 } \cos \alpha_m \cos \theta + {\sqrt{3} \over 2} \sin \theta \\
\sin \alpha_m \cos \theta
\end{array}\right).
\end{equation}

To determine the dependence of the phonon motion on the polarization angle $\theta$, we can use Eq.~\ref{eq:F} which gives the time-dependent force on an atom along one of the phonon eigenvectors.  The resulting motion is then given by the classical equation of motion
\begin{equation}
\frac{d^2u_j}{dt^2} + \Omega_j^2 u_j + 2 \gamma_j \frac{d u_j}{dt} = F_j(t)/m\label{eq:q_ODE}
\end{equation}
where $m$ is the mass of an atom, $\Omega_j$ is the mode frequency, $\gamma_j$ is the damping rate, and $u_j$ is the position of the atom along the eigenvector.  The initial conditions at $t=-\infty$ are $u_i = \frac{d u_j}{d t} = 0$.  

We first consider the $A_{1g}$ mode, where the relevant coordinate is $u_z$.  The behavior of this mode is usually described using the ``displacive excitation'' model~\cite{Zeiger:1992tp,Kuznetsov:1994jv}.  The idea here is that electronic excitation causes a sudden displacement  of the quasi-equilibrium value of the phonon coordinate away from zero, which in Eq.~\ref{eq:q_ODE} is equivalent to a step-like behavior in $F_z(t)$.    If the lifetime of the electronic relaxation is much longer than that of a phonon period, the solution of Eq.~\ref{eq:q_ODE} for $u_z$ is a cosine-like oscillation about the displaced quasi-equilibrium value of the coordinate.  The dependence of the $u_z$ motion on the pump polarization in the displacive excitation model is given simply by the dependence of the absorption of the pump light on the polarization.  As noted above, in bismuth at 800~nm this dependence is relatively weak and leads to overall variation in the pump-induced displacement of approximately 10\%.

We now consider the $E_g$ modes.  In this case the modes are driven by electronic excitation to states that break some symmetry operations of the crystal, a subset of the states that form the conduction bands of bismuth.  The lifetime of these states is then a critical parameter for characterizing the structural response of the system.  If these states are long-lived, Stevens et al. argue that a similar argument would apply as for the $A_{1g}$ mode and yield a displacive form of the force on the $E_g$ coordinates and a cosine-like displaced oscillation.  This is, however, clearly not the case in the experiment:  the $u_y$ coordinate oscillates about zero, and has a phase closer to that of a damped sine function.  Riffe and Sabbah\cite{Riffe:2007wc} have proposed an extension of the model of Stevens et al. that can cover the case where the relevant electronic state lifetime is very small, as might be expected from strong carrier-carrier scattering.     
The effective force on the $E_g$ coordinate is then approximately proportional to the instantaneous intensity of the pump pulse inside the crystal.  In the frequency-space formulation of the force given in Eq.~\ref{eq:F}, this is equivalent to approximating the elements $d$ and $f$ of $\pi_{jk}^{x,y}(\omega,\omega-\Omega)$ as independent of $\Omega$ for values of $\Omega$ comparable to or smaller than the inverse pump pulse duration.
Applying Eq.~\ref{eq:F} for the force along the $u_y$ coordinate and solving for the motion of $u_y$ gives
\begin{equation}
u_y(t) = \int_{-\infty}^{\infty} g(t-t^\prime) h(t^\prime) dt^\prime,
\end{equation}
where 
\begin{equation}
g(t) = \frac{1} {\tau \sqrt{\pi} } e^{-t^2/\tau^2}
\end{equation}
is the area-normalized intensity envelope of the laser pulse, 
\begin{equation}
h(t) = \left\{\begin{array}{ll}
0 & t<0\\
u_y^{(0)} e^{-\gamma_y t}
\sin \sqrt{\Omega_y^2 - \gamma_y^2}t & t\ge 0  \end{array}\right.,
\end{equation}
\begin{equation}
u_y^{(0)} = {  \mathcal{F} v_c \over 4 m c   \sqrt{\Omega_y^2-\gamma_y^2}} {1 \over |1+\sqrt{\epsilon}|^2} 
\left[
C + D \cos (2 \theta - \delta)
\right], \label{eq:u0}
\end{equation}
\begin{equation}
C = A - 4 d,
\end{equation}
\begin{equation}
D = \sqrt{A^2+B^2},
\end{equation}
\begin{equation}
\delta = \tan^{-1} \left({B \over A}\right),
\end{equation}
\begin{equation}
A = d (3 + \cos 2\alpha_m) +  2 f \sin 2 \alpha_m,
\end{equation}
\begin{equation}
B = 4 \sqrt{3} {\left(- d \cos \alpha_m + f \sin \alpha_m \right)}
\end{equation}
and $\mathcal{F}$ is the incident laser fluence.

\begin{figure}
\includegraphics[width=15cm]{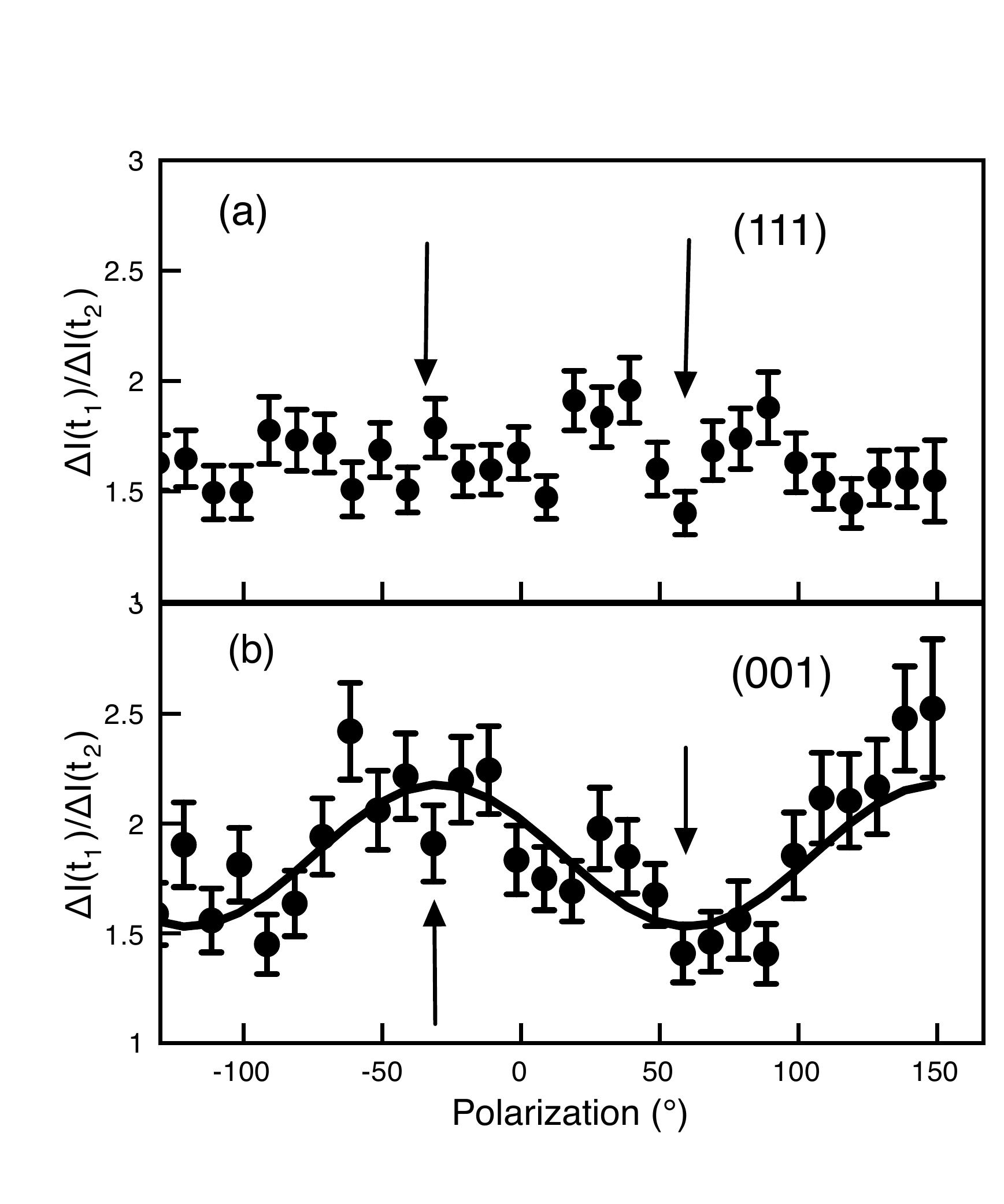}
\caption{Polarization dependence of the ratio of the pump-induced changes between times $t_1=80\,\textrm{fs}$ and $t_2=370\,\textrm{fs}$ after excitation, for (a) the $(111)$ reflection and (b) the $(001)$ reflection.  While diffraction from $(111)$ shows no dependence on the polarization, that from the $(001)$ planes shows a strong dependence with a period of $180^\circ$.  The solid curve shows a fit to a model discussed in the text.
}\label{fig:poldep}
\end{figure}

\begin{figure}
\includegraphics[width=15cm]{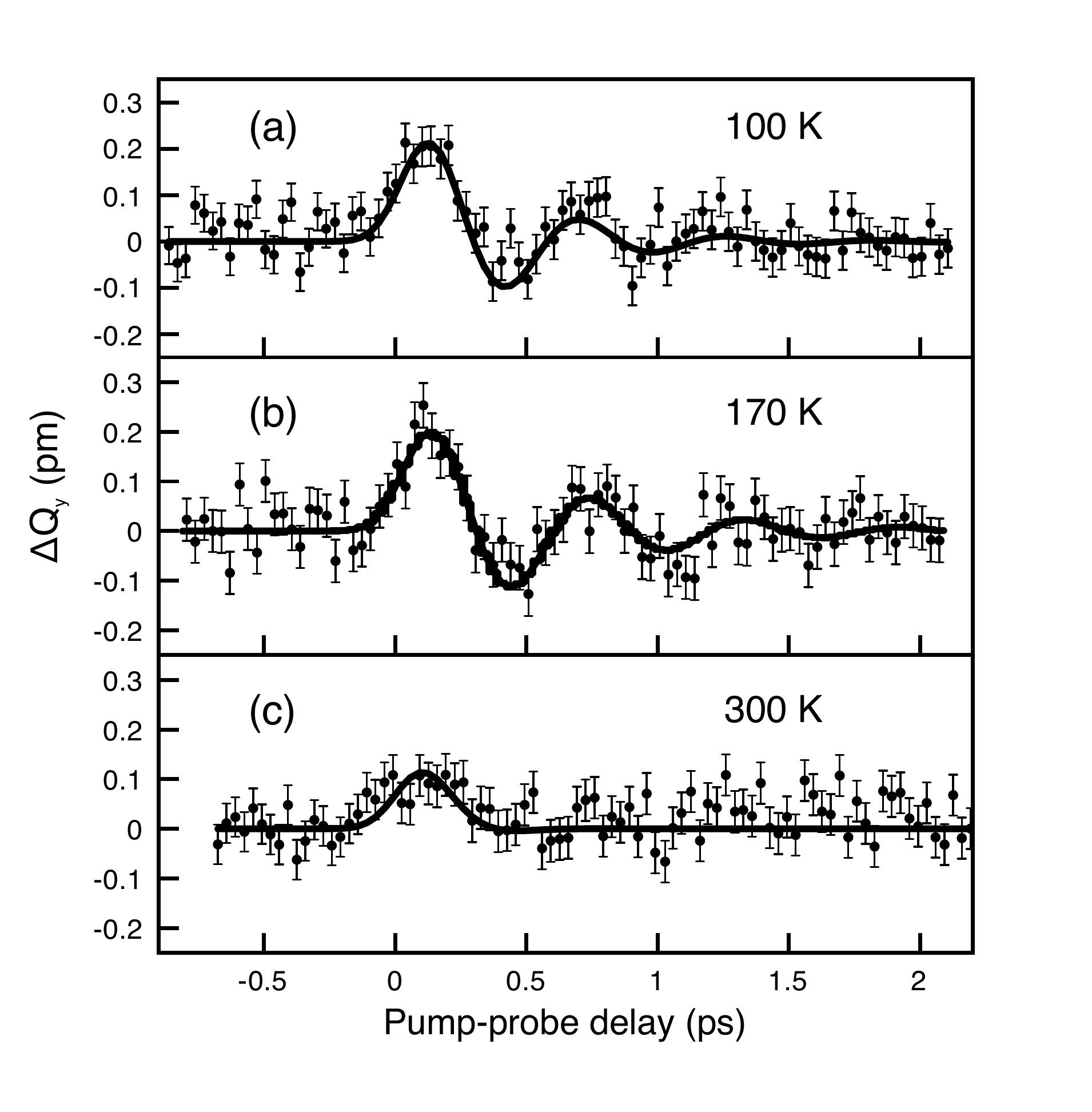}
\caption{Difference in the $u_y$ coordinate as a function of pump-probe delay time for excitation polarizations at $\theta = -30^\circ$ and $\theta = 60^\circ$.  The different panels show data taken at sample temperatures of (a) 100 K, (b) 170 K and (c) 300 K.  The solid lines are fits to a damped coherent $E_g$ mode discussed in the text.}\label{fig:duT}
\end{figure}

The amplitude of the displacement is the sum of a term that is polarization independent and a term that varies as $\cos(2\theta-\delta)$. 
 The polarization dependence of the time-resolved diffraction from $(001)$ shown in Fig.~\ref{fig:poldep}(b) shows a $\cos(2\theta-\delta)$ dependence in $\Delta I(t_1)/\Delta I(t_2)$. 
The solid line here shows a fit to the form of Eq.~\ref{eq:u0}.   The fitted value of $\delta = -1.04 \pm 0.14$ gives us directly an estimate for the ratio $f/d = -0.07\pm 0.14$.  
Figure~\ref{fig:duT} shows the {difference} in the $u_y$ dynamics between the polarizations $\theta = -30^\circ$ and $\theta = 60^\circ$, inferred from the diffraction data using Eq.~\ref{eq:I001}.  
The solid lines in Fig.~\ref{fig:duT} are fits to $\Delta u_y(t) = u_y(t,\theta=-30^\circ) - u_y(t,\theta=60^\circ)$ using the above expressions.  Table 1 summarizes the resulting fit parameters.

Our estimates of $\delta$ and $D$ allow us to estimate absolute values for the Raman tensor components $d$ and $f$ under the assumption that $\delta$ is independent of temperature.  These are listed in table 1.  These values imply that the order of magnitude of $10^{-2}$ for changes in the dielectric constant due to the phonon motion, which is roughly consistent with the magnitude of reflectivity changes observed in optical reflectivity measurements~\cite{Hase:2002in,Misochko:2006gd,Li:2011vb}.  The large decrease in signal when increasing temperature from 170~K to 300~K appears to be due to a large increase in the damping of the mode, suggesting that the mechanism of damping is related to interactions with incoherent phonon modes that are more populated at the higher temperature.

We also note that the peak-to-peak amplitude of the E$_g$ oscillations (0.2 pm) is more than a factor of 10 smaller than the atomic motions associated with the A$_{1g}$ mode ($2.7 \pm 0.1$ pm) at the measured excitation fluence.  The difference may be understood by considering once again the lifetime of the electronic excitations that drive each mode.  We can estimate the relative force on the excited phonon by taking the Fourier transform of Eq.~\ref{eq:F} and evaluating its magnitude at the phonon frequency.  Assuming states with a linear decay constant $\gamma$ contribute to the phonon force is (for short driving pulses)
\begin{equation}
F_j(\Omega) \propto {1 \over \gamma + i \Omega_j}.
\end{equation}
For the A$_{1g}$ mode $\gamma \ll \Omega_z$, and so the amplitude of the phonon is approximately proportional to $1/\Omega_z$.  The short-lived electronic states that drive the E$_g$ modes, however, have a much higher damping $\gamma \gg \Omega_{x,y}$.  The amplitude of the phonon is then proportional to $1/\gamma$.  If we assume the proportionality constants are of the same order of magnitude, from the relative size of the phonon amplitudes we can estimate $1/\gamma \sim 4\,\textrm{fs}$.  This is consistent with our expectations based on the sinusoidal phase of the $E_g$ oscillations.  Note that the sinusoidal phase would also be expected for impulsive Raman scattering from coupling to virtual electronic transitions, which gives an additional purely real contribution to $F(\Omega)$.  The spectral dependence of the relative Raman cross section indicates, however, that the resonant contributions dominate in this range of excitation wavelengths~\cite{Renucci:1973uj}.

The measured magnitude of the coherent E$_g$ modes allows us to test proposed models of  phonon-phonon coupling in bismuth.  In Ref.~\onlinecite{Hase:2002in}, Hase et al. report transient reflectivity measurements on bismuth that show possible evidence of coupling between the A$_{1g}$ and E$_g$ modes at excitation levels approximately equivalent to those used in our diffraction experiment\footnote{We make this comparison based on a value of 2.65~THz for the A$_{1g}$ frequency from a fit to the $(111)$ diffraction, using a model  described in the caption of Fig. 2 of Ref.~\onlinecite{Johnson:2008ku}.  This can be compared against a value of 2.67~THz at a nominal fluence of 7.6 mJ/cm$^2$ studied in Ref.~\onlinecite{Hase:2002in}.  Since the A$_{1g}$ frequency is strongly fluence dependent, it serves as a sensitive measure of the true excitation level.}.    In this work a low-frequency side-peak in the spectral density of the transient reflectivity appears with an amplitude of approximately 6\% that of the main A$_{1g}$ peak.
In Ref.~\onlinecite{Zijlstra:2006wt}, Zijlstra et al. propose to explain this side-peak as a consequence of coupling between the A$_{1g}$ and E$_g$ modes.  They performed numerical simulations to support this idea.  The simulated data reproduce a side-peak at a similar frequency as seen in the optical reflectivity data, but the magnitude of the peak depends strongly on the assumed coherent E$_g$ amplitude. 
For a peak-to-peak E$_g$ amplitude of $0.15  \Delta u_z$, where $\Delta u_z$ is the initial peak-to-peak amplitude of the A$_{1g}$ mode, the phonon-phonon coupling model predicts a spectral side-peak with a magnitude only  $7\times 10^{-7}$ that of the dominant A$_{1g}$ peak.  Using our measured $\Delta u_z = 2.7\,\textrm{pm}$, we see that the peak-to-peak E$_g$ amplitude is in reality $0.075 \Delta u_z$, a factor of two smaller.  We conclude that phonon-phonon coupling as presented in Ref.~\onlinecite{Zijlstra:2006wt} is not sufficient to explain this feature of the optical reflectivity data.  

These results also confirm the interpretation of Ref.~\onlinecite{Johnson:2009ty}, where the oscillatory dynamics of diffraction from the $(1\,0\,\bar{1})$ and $(1\,1\,\bar{2})$ planes in bismuth were  ascribed to excitation-induced changes in the Debye-Waller factor.  Contributions from the E$_g$ modes were excluded based on the time dependence of the observed dynamics, which were inconsistent with the expectations from coherent E$_g$ oscillations.  The present work allows us to also make an upper bound on the magnitude of coherent E$_g$ contributions of $8 \times 10^{-6}$ for the $(1\,\bar{1}\,0)$ peak and $3 \times 10^{-5}$ for the $(1\,1\,\bar{2})$ peak.   These are much smaller than the $10^{-3}$ noise level on these measurements, implying that the observed dynamics are indeed related to changes in the Debye-Waller factor and not coherent E$_g$ phonons.   A small contribution of the approximately 1.4~THz Debye-Waller oscillations may also explain the low-frequency side-peak seen in the optical data of ref.~\onlinecite{Hase:2002in}. Alternatively, the side-peak could also be explained as a direct contribution to the reflectivity from the coherent E$_g$ mode because of a small optical alignment error.

As discussed in the introduction, coherent control over non-fully-symmetric optical phonons may be a viable route to the control of a variety of phase transitions.  For this purpose much larger amplitudes of coherent excitation are desirable.  One key to increasing the amplitude of the coherent motion under the resonant Raman scheme is to increase the lifetime of the electronic states that drive the transition. 
Greater amplitudes may be realized in semiconducting or insulating systems pumped just above bandgap, or in strongly correlated electron systems where the final electronic states are either long-lived or take significantly longer times to equilibrate.  Alternatively, direct dipole excitation using coherent mid-infrared or terahertz frequency pulses show promise in this area for cases where the mode is infrared active.

\begin{acknowledgments}
These experiments were performed on the X05LA beamline at the Swiss Light Source, Paul Scherrer Institut, Villigen, Switzerland. 
This work was supported by the NCCR Molecular Ultrafast Science and Technology (NCCR-MUST), a
research instrument of the Swiss National Science Foundation (SNSF).  
\end{acknowledgments}

\clearpage

\clearpage

\clearpage

\clearpage

\clearpage

\clearpage
\begin{table}
\begin{tabular*}{0.9\textwidth}{@{\extracolsep{\fill}}cccccc}
\hline
Temperature (K) & $\nu_y$ (THz) & $\gamma_y$ (ps$^{-1}$) & $D$ (pm$^{-1}$) & d (pm$^{-1}$) & f (pm$^{-1}$)\\
\hline
100 & $1.84 \pm 0.06$  & $2.9 \pm 0.5$ & $10.5 \pm 1.3$ & $2.2 \pm 0.4$   & $-0.2 \pm 0.3$ \\
170 & $1.8 \pm 0.04$   & $1.8 \pm 0.3$ & $8.4 \pm 1.2$  & $1.7 \pm 0.3$   & $-0.1 \pm 0.3$ \\
300 & $1.9 \pm 0.7$    & $12 \pm 10$   & $9 \pm 7$      & $1.8 \pm 1.1$   & $-0.1 \pm 1.0$ \\
\hline
\end{tabular*}
\caption{Summary of fit parameters for the data in Fig.~\ref{fig:duT}, using the fit function discussed in the text.  Here $\nu_y = \Omega_y/2\pi$.    
The estimated uncertainties reported for $D$, $d$ and $f$ do not take into account an additional 10\% uncertainty in the excitation fluence, stemming from the uncertainty in the laser spot size that was the same for all measurements.}
\end{table}
\clearpage

\bibliography{frompapers}

\end{document}